\documentclass{IEEEtran4PSCC}

\usepackage[pdftex]{graphicx}
\usepackage[cmex10]{amsmath}

\usepackage[draft]{minted}
\usepackage{float}
\floatstyle{ruled}
\newfloat{code}{thp}{lop}
\floatname{code}{Code Block}

\usepackage{booktabs}

\begin{document}

\title{PowerModelsDistribution.jl: An Open-Source Framework for Exploring Distribution Power Flow Formulations}

\author{\IEEEauthorblockN{David M Fobes\IEEEauthorrefmark{1},
Sander Claeys\IEEEauthorrefmark{2},
Frederik Geth\IEEEauthorrefmark{3}, and
Carleton Coffrin\IEEEauthorrefmark{1}
}
\IEEEauthorblockA{\IEEEauthorrefmark{1} Los Alamos National Laboratory (LANL)\\
Los Alamos, New Mexico, USA\\
\{dfobes,cjc\}@lanl.gov}
\IEEEauthorblockA{\IEEEauthorrefmark{2} Katholieke Universiteit Leuven (KU Leuven)\\
Leuven, Belgium\\
sander.claeys@kuleuven.be},
\IEEEauthorblockA{\IEEEauthorrefmark{3} Commonwealth Scientific and Industrial Research Organisation (CSIRO)\\
Canberra, Australia\\
frederik.geth@csiro.au}
}

\maketitle
\begin{abstract}
    In this work we introduce PowerModelsDistribution, a free, open-source toolkit for distribution power network optimization, whose primary focus is establishing a baseline implementation of steady-state multi-conductor unbalanced distribution network optimization problems, which includes implementations of Power Flow and Optimal Power Flow problem types. Currently implemented power flow formulations for these problem types include AC (polar and rectangular), a second-order conic relaxation of the Branch Flow Model (BFM) and Bus Injection Model (BIM), a semi-definite relaxation of BFM, and several linear approximations, such as the simplified unbalanced BFM. The results of AC power flow have been validated against OpenDSS, an open-source ``electric power distribution system simulator'', using IEEE distribution test feeders (13, 34, 123 bus and LVTestCase), all parsed using a built-in OpenDSS parser. This includes support for standard distribution system components as well as novel resource models such as generic energy storage (multi-period) and photovoltaic systems, with the intention to add support for additional components in the future.
\end{abstract}


\section{Introduction}
\subsection{Background}
The past few decades has seen a remarkable increase in the penetration of distributed energy resources (DER), such as photovoltaic systems, wind turbines, storage devices, and fuel cells, as well as the unique demands of controllable loads, such as electric vehicle chargers and heating, ventilation and air conditioning (HVAC) systems, within electric distribution networks. While the application of DERs can mitigate the need for more traditional transmission network expansion, \emph{i.e.}, via the addition of generators, the operation and control of distribution networks remains a challenge. In recent years this has driven considerable interest in the research community to develop multi-conductor power network optimization formulations for various applications, such as a networked micro-grid design. While the number of mathematical formulations for distribution system modeling has increased, few open-source tools are yet available, and none that enable rapid development of the newest formulations and optimization problems, to the best of our knowledge.

\subsection{The Development of PowerModels}
Recently, in response to an explosion of the number of power flow approximations and relaxations appearing in the literature for transmission networks, PowerModels \cite{Coffrin2017} was offered as a free, open-source toolkit for the optimization of steady-state power transmission networks. Written in Julia, a high-level high-performance programming language for numerical computing, and utilizing JuMP \cite{Dunning2015}, PowerModels provides a powerful expansive modeling layer for optimization; PowerModels is engineered to decouple problem specifications, \emph{e.g.}, Optimal Power Flow (OPF) or Optimal Transmission Switching (OTS), from formulations, \emph{e.g.}, AC or second-order cone (SOC) relaxations. This decoupled design allows for more accurate, faster, and more detailed comparisons between the variety of problem specifications and formulations that are constantly emerging from the research community.

PowerModels, being easily extensible, has generated a number of spin-off packages, for modeling, \emph{e.g.}, DC networks and HVDC converters \cite{Ergun2019}, maximum load delivery under contingencies \cite{Coffrin2019}, or the impact of geomagnetic disturbances\footnote{https://github.com/lanl-ansi/PowerModelsGMD.jl}.

\subsection{Introducing PowerModelsDistribution}
Building on the success of PowerModels in the area of \emph{transmission} power network optimization, in this work we introduce PowerModelsDistribution, a free, open-source toolkit for \emph{distribution} power network optimization developed on top of JuMP, a mathematical programming abstraction layer for optimization. This project currently focuses on establishing a baseline implementation of steady-state multi-conductor unbalanced distribution network optimization problems. Many recently published formulations (for an overview, see \cite{Molzahn2017a}) are already included, from non-convex nonlinear forms, to convex relaxations, to linear approximations of the unbalanced power flow equations. Although both approximations and relaxations might not satisfy the originating formulation, e.g. non-linear power flow equations, unlike approximations, relaxations can provide information about the original problem. For example, when a relaxation is infeasible, it certifies that the original problem is also infeasible. Furthermore, relaxations can also provides a lower bound on the objective, which can be useful for global optimization methods \cite{Molzahn2017a}. Recent work extends relaxations of the power flow equations to multi-phase networks \cite{Molzahn2014a,Shen2019,Zhao2017a}.

PowerModelsDistribution is a tool designed to build and compare unbalanced power network \emph{optimization} problems, \emph{e.g.}, Optimal Power Flow. In this work, to demonstrate the accuracy of our mathematical formulations, we focus on the results of power flow feasibility; the numerical results of the AC power flow are validated w.r.t. OpenDSS on IEEE distribution test cases \cite{Mather2017}, which include a broad range of distribution network components. While we support the OpenDSS data format as input, it is not the intention of this package to replicate the features of OpenDSS, but to leverage their existing mature data format for building network cases. Within its robust OpenDSS parser, PowerModelsDistribution supports a subset of DER, including generic energy storage and photovoltaic systems, with support for additional DER components coming in the future. In time, we hope that PowerModelsDistribution will emerge as an essential part of the distribution network \emph{optimization} and analysis toolkit.

This paper therefore explores the design of an open-source toolkit for simulation and optimization of distribution network power flows, where careful consideration of phase unbalance effects is important. Modeling focuses on (quasi-)steady state physics, described in the frequency domain.

\section{Review of Distribution modeling Approaches} \label{sec_lit_review}
For a recent, in-depth discussion of free, open-source power system simulation and optimization tools we refer to \cite{Meinecke2018}; here we focus on the subset of tools that are capable of representing the effects of unbalanced phases.
Phase unbalance is the consequence of unbalanced loading, such as single-phase loads or generators connected to a three-phase network, or unequal branch impedance, \emph{e.g.}, caused by a lack of transposition of conductors with a non-equilateral geometry.

Calculation of the power flow in networks featuring non-negligible phase unbalance has long been a topic of interest \cite{Berg1967}. Several open-source tools include algorithms for solving unbalanced power flows: OpenDSS \cite{Dugan2011}, originally developed at the Electrotek Concepts in 1997, and open-sourced at the Electric Power Research Institute in 2008, was one of the first free tools for distribution system simulation; in 2003, the Pacific Northwest National Laboratory started development of Gridlab-D \cite{Chassin2008}; recently, PandaPower \cite{Meinecke2018} and GridCal \cite{Vera2019} have added support for unbalanced power flow.

A variety of commercial options is also available and in use globally by distribution system operators for solving power flows, including PowerFactory, PSS/Sincal, Neplan, and CYME. While most of these rely on proprietary file formats, several conversion tools have been developed to enable data exchange:
\begin{itemize}
    \item DiTTo\footnote{https://github.com/NREL/ditto}, supports conversion between a large variety of open-source and commercial tools;
    \item GRIDAPPSD/Powergrid-Models\footnote{https://github.com/GRIDAPPSD/Powergrid-Models} provides CIM interfaces for GridLAB-D and OpenDSS.
\end{itemize}
Fortunately, standardization of data models for distribution system management (CIM Common Distribution Power System Model) is being developed as part of IEC 61968 \cite{IEC61968}.

\subsection{Foundations of Distribution Network Modeling}
The physics of power flow in networks with phase unbalance are described in the framework of Kirchhoff's circuit laws; the equations are linear in current and voltage for fixed impedances. In the context of modeling the coupling between conductors, the impedance becomes a matrix parameter composed of self-impedance (diagonal) and mutual impedances (off-diagonals). Therefore, the current-voltage variable space represents a natural choice for distribution modeling.

\begin{figure}[tbh]
    \centering
    \includegraphics[width=0.95\columnwidth]{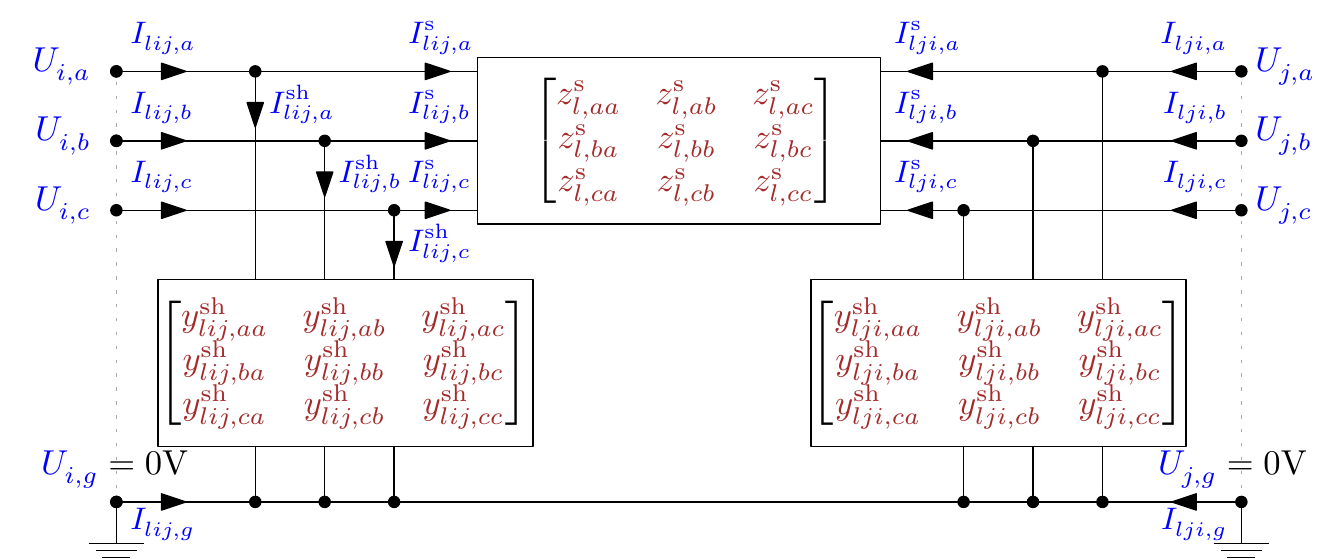}
    \caption{Unbalanced $3\times3$ $\Pi$-model branch in voltage and current variables. }  \label{fig_linemodel_iv_3x3}
\end{figure}

The size of the impedance matrix varies on the context; three-wire and Kron-reduced four-wire networks use $3\times3$ impedance matrices, but generalizations that include an explicit neutral conductor voltage and current variables, or even earth voltage rise effects \cite{Ciric2003}, can result in up to $5\times 5$ impedance matrices. In practice, in the collection of data and the construction of the mathematical models a variety of approximations are made \cite{Urquhart}, including:
\begin{itemize}
    \item neglecting branch shunt admittance,
    \item assuming constant-power load behavior,
    \item assuming perfect grounding of the neutral at all buses,
    \item assuming loads are balanced over the phases.
\end{itemize}
To support nontrivial edge cases, PowerModelsDistribution has chosen generic branch and bus representations; formulations are standardized on multi-conductor $\Pi$ sections, supporting full matrices for the series and shunt elements (variable size). Furthermore, the admittance shunts on either side of the conductor are not necessarily identical, thereby enabling $\Gamma$ sections. The branch model of PowerModelsDistribution is illustrated in Fig. \ref{fig_linemodel_iv_3x3}.

In the following, we illustrate the primary features of the physics of unbalanced power flow in the current-voltage variable space, assuming constant impedances.

\subsubsection{Voltage and Current Variables}
The bus voltage vector $\mathbf{U}_i$ stacks the scalar complex-value voltages for each phase,
\begin{IEEEeqnarray}{C}
    \mathbf{U}_i =
    \begin{bmatrix}
        U_{i,a}    \\
        U_{i,b}     \\
        U_{i,c}
    \end {bmatrix}
        =   \begin{bmatrix}
        U_{i,a}^{\text{re}}    \\
        U_{i,b}^{\text{re}}     \\
        U_{i,c}^{\text{re}}
    \end {bmatrix} + j
    \begin{bmatrix}
        U_{i,a}^{\text{im}}    \\
        U_{i,b}^{\text{im}}     \\
        U_{i,c}^{\text{im}}
    \end {bmatrix}.
\end{IEEEeqnarray}
Similarly, series and shunt current vectors are defined:
\begin{IEEEeqnarray}{C}
    \mathbf{I}_{lij} =
    \begin{bmatrix}
        I_{lij,a}    \\
        I_{lij,b}     \\
        I_{lij,c}
    \end {bmatrix} =
    \begin{bmatrix}
        I_{lij,a}^{\text{s}}    \\
        I_{lij,b}^{\text{s}}     \\
        I_{lij,c}^{\text{s}}
    \end {bmatrix} +
    \begin{bmatrix}
        I_{lij,a}^{\text{sh}}    \\
        I_{lij,b}^{\text{sh}}     \\
        I_{lij,c}^{\text{sh}}
    \end {bmatrix}
    =\mathbf{I}^{\text{s}}_{lij} + \mathbf{I}^{\text{sh}}_{lij}. \label{eq_currents}
\end{IEEEeqnarray}

\subsubsection{Ohm's Law}
The voltage at node $j$ w.r.t. node $i$, and the drop over the series impedance $\mathbf{Z}_l$ in a branch $l$ is,
\begin{IEEEeqnarray}{C}
    \mathbf{U}_j = \mathbf{U}_i - \mathbf{Z}_l \mathbf{I}^{\text{s}}_{lij}.  \label{eq_ohms}
\end{IEEEeqnarray}
The branch shunt admittance $\mathbf{Y}^{\text{sh}}_{lij}$ draws a current,
\begin{IEEEeqnarray}{C}
    \mathbf{I}^{\text{sh}}_{lij}  = \mathbf{Y}^{\text{sh}}_{lij} \mathbf{U}_i. \label{eq_admittance}
\end{IEEEeqnarray}

\subsubsection{Kirchhoff's Current Law}

We can substitute \eqref{eq_ohms} and \eqref{eq_admittance} into \eqref{eq_currents} to obtain the total current as a function of voltage, with $\mathbf{Y}^{\text{s}}_{l} = (\mathbf{Z}^{\text{s}}_{l} )^{-1}$:
\begin{IEEEeqnarray}{C}
    \mathbf{I}_{lij}  =  \mathbf{Y}^{\text{sh}}_{lij} \mathbf{U}_i  +  \mathbf{Y}^{\text{s}}_{l} (\mathbf{U}_i  - \mathbf{U}_j). \label{eq_total_current_voltage}
\end{IEEEeqnarray}

At buses, current flows between the branches $lij \in \mathcal{T}$, bus shunts $b \in \mathcal{B}$, loads $d  \in \mathcal{D}$ and generators $g \in \mathcal{G}$ connected,
\begin{IEEEeqnarray}{C}
    \sum_{lij \in \mathcal{T}(i)} \mathbf{I}_{lij}  =  \sum_{b \in \mathcal{B}(i)} \mathbf{I}^{\text{sh}}_{b} +  \sum_{g \in \mathcal{G}(i)} \mathbf{I}_{g} -  \sum_{d \in \in \mathcal{D}(i)}  \mathbf{I}_{d}. \label{kcl_current}
\end{IEEEeqnarray}

\subsubsection{Complex Power Flow and Sources/Sinks}
The power flow into a branch is defined
\begin{IEEEeqnarray}{C}
    \mathbf{S}_{lij} = \mathbf{U}_i ( \mathbf{I}_{lij})^{\text{H}}, \label{eq_power_line_def}
\end{IEEEeqnarray}
Note that this equation is nonlinear and non-convex, \emph{i.e.}, the product of the voltage vector at the bus is multiplied by the conjugate transpose of the current vector. This results in $\mathbf{S}_{lij}$ being a matrix, whose diagonal contains the elements $S_{lij,pp} = U_{i,p} (I_{lij,p})^{*} \,\,\, \forall p \in \{a,b,c\}$.
Analogously, we define the power of loads and generators,
\begin{IEEEeqnarray}{C}
    \mathbf{S}_{d} = \mathbf{U}_i ( \mathbf{I}_{d})^{\text{H}} \\
    \mathbf{S}_{g} = \mathbf{U}_i ( \mathbf{I}_{g})^{\text{H}}.
\end{IEEEeqnarray}
If  generators are dispatchable, we define bounds on $diag( \mathbf{S}_{g})$.

\subsubsection{Matrix, Vector and Scalar Forms} \label{sec_matrix_vector_scalar}
Next, we substitute \eqref{eq_total_current_voltage} in into \eqref{eq_power_line_def}, obtaining the \emph{matrix} equality,
\begin{IEEEeqnarray}{C}
    \mathbf{S}_{lij}  = \mathbf{U}_i  ( \mathbf{U}_i)^{\text{H}}( \mathbf{Y}^{\text{sh}}_{lij})^{\text{H}} +   \mathbf{U}_i (\mathbf{U}_i  - \mathbf{U}_j)^{\text{H}}( \mathbf{Y}^{\text{s}}_{lij})^{\text{H}}. \label{eq_total_power_voltage}
\end{IEEEeqnarray}
Although in general the off-diagonal terms in \eqref{eq_total_power_voltage} are redundant (\emph{i.e.,} rotated/scaled versions of the diagonal terms), when performing lifting and convex relaxation, these off-diagonal terms are not necessarily redundant. Dropping the off-diagonal terms results in a \emph{vector} form. Finally, the diagonal terms are \emph{scalarized} and converted to reals before implementation, \emph{e.g.}, for the active power flow, parameterized on the diagonal elements $p$ by:
\begin{multline}{}
    P_{lij,pp} =
    \sum_{q \in \mathcal{P}} ( U^{\text{re}}_{i,p}   U^{\text{re}}_{i,q}   + U^{\text{im}}_{i,p}     U^{\text{im}}_{i,q}  )  \left( g^{\text{s}}_{l,pq} + g^{\text{sh}}_{lij,pq} \right)  \\
    +  \sum_{q \in \mathcal{P}} ( U^{\text{im}}_{i,p}   U^{\text{re}}_{i,q}   - U^{\text{re}}_{i,p}     U^{\text{im}}_{i,q}  )  \left( b^{\text{s}}_{l,pq} + b^{\text{sh}}_{lij,pq} \right)   \\
    -     \sum_{q \in \mathcal{P}}   ( U^{\text{re}}_{i,p}   U^{\text{re}}_{j,q}   + U^{\text{im}}_{i,p}     U^{\text{im}}_{j,q}  )   g^{\text{s}}_{l,pq}    \\
    -     \sum_{q \in \mathcal{P}} ( U^{\text{im}}_{i,p}   U^{\text{re}}_{j,q}   + U^{\text{re}}_{i,p}     U^{\text{im}}_{j,q}  )   b^{\text{s}}_{l,pq}    .  \label{eq_active_rectangular_pf}
\end{multline}
Note that alternatively, this equation could be derived using polar voltage variables instead. These alternatives can have different numerical properties, and different relaxations can be obtained when starting from different forms.

\subsection{Separating Modeling and Solving}
Now understanding the physics-based foundations, we ask whether distribution system models can be stated as mathematical optimization programs, referred to henceforth simply as mathematical programs. Mathematical modeling tools allow for a clear separation of the model and the solution method, which results in a declarative programming approach, as the modeler merely describes the problem structure, and requests the solver to return a solution satisfying a set of conditions without prescribing \emph{how} the solution should be found. Well-known mathematical modeling layers include: GAMS, AMPL, AIMMS, Yalmip, JuMP, Pyomo,  CVXPy, and Convex. These modeling layers access the different solution algorithms, \emph{e.g.}, solvers for nonlinear programming (NLP), semidefinite programming (SDP), second-order cone programming (SOCP), or linear programming (LP) problems, which allows for the ability to switch solution approaches quickly without having to re-implement the equations. This is an essential feature for the rapid prototyping that is central to algorithmic research, allowing for direct comparison of solver performance on identical problems and formulations.

Such an approach has not been commonly used in distribution system modeling; the bulk of the work to date on distribution modeling has focused on power flow \emph{algorithms} describing processes for computing solutions to specific problems (\emph{e.g.}, unbalanced power flow), as opposed to optimization problems on which we focus on here. When it comes to solving power flow problems, it is an open question whether such mathematical programs for optimization can reliably produce results with similar accuracy and modeling fidelity as state-of-the-art distribution power flow tools such as GridLab-D and OpenDSS. Due to this approach, the performance of the power flow solve is much less than that achieved by dedicated power flow solvers by design, and is not meant as a replacement for dedicated power flow, but to serve as a means of fidelity comparison between other tools. In PowerModelsDistribution, the power flow is obtained via the same interface as for example OPF, utilizing constrained optimization solvers.

To facilitate collaboration, mathematical modeling layers should enable swift implementation and maintainability. Features that are important in this context are:
\begin{itemize}
    \item supporting vector / matrix equations (automatic scalarization),
    \item automatic derivative computations of nonlinear equations, and
    \item support for semidefinite variables and/or constraints,
\end{itemize}
all of which are supported by the Julia-based mathematical programming package JuMP \cite{Dunning2015}.

\section{Distribution System Mathematical Programs} \label{sec_math_programs}
The following section provides an overview of the components and modeling assumptions of PowerModelsDistribution and describes several approaches to the mathematical formulation of distribution network physics.

\subsection{Component Models}
In distribution networks, a broad variety of power transfer and consumption/generation sources are encountered. This section details which components are prioritized, and which mathematical modeling aspects are considered.

\subsubsection{Buses and nodes (\texttt{bus})}
Buses have multiple nodes, \emph{i.e.}, one for each conductor. In optimization we want to optimally operate the network such that voltage limits are satisfied. The voltage magnitudes are thus given lower and upper bounds,
\begin{IEEEeqnarray}{C}
    \begin{bmatrix}
        U_{i,a}^{\text{min}}    \\
        U_{i,b}^{\text{min}}     \\
        U_{i,c}^{\text{min}}
    \end {bmatrix} \leq
    \begin{bmatrix}
        |U_{i,a}|    \\
        |U_{i,b}|     \\
        |U_{i,c}|
    \end {bmatrix} \leq
    \begin{bmatrix}
        U_{i,a}^{\text{max}}    \\
        U_{i,b}^{\text{max}}     \\
        U_{i,c}^{\text{max}}
    \end {bmatrix},
 \end{IEEEeqnarray}
typically chosen in accordance with grid codes. %such as EN 50160.
Note that bounds can also be applied to the sequence components. For instance, the ratio of the negative to positive sequence components of the voltage phasor should be below 2\% \cite{Girigoudar}.

\subsubsection{Cables and overhead lines (\texttt{branch})}
Both cables and overhead lines are represented through multi-conductor $\Pi$ sections (Fig. \ref{fig_linemodel_iv_3x3}). Flow bounds, on current or power magnitude, are applied to both ends of a $\Pi$-section,
\begin{IEEEeqnarray}{C}
    \begin{bmatrix}
        |I_{lij,a}|    \\
        |I_{lij,b}|     \\
        |I_{lij,c}|
    \end {bmatrix} \leq
    \begin{bmatrix}
        I_{lij,a}^{\text{max}}    \\
        I_{lij,b}^{\text{max}}     \\
        I_{lij,c}^{\text{max}}
    \end {bmatrix} ,
    \begin{bmatrix}
        |S_{lij,aa}|    \\
        |S_{lij,bb}|     \\
        |S_{lij,cc}|
    \end {bmatrix} \leq
    \begin{bmatrix}
        S_{lij,aa}^{\text{max}}    \\
        S_{lij,bb}^{\text{max}}     \\
        S_{lij,cc}^{\text{max}}
    \end {bmatrix} .
\end{IEEEeqnarray}
The $\Pi$ sections are easily generalized for a different number of conductors.
Reductions in number of conductors (\emph{e.g.}, a single-phase branch off the feeder) are also supported.

\subsubsection{Bus shunts (\texttt{shunt})}
Bus shunts $b \in \mathcal{B}$, defined through an admittance $\mathbf{Y}^{\text{sh}}_{b}$, can be used to represent components such as capacitor banks,
\begin{IEEEeqnarray}{C}
    \mathbf{I}^{\text{sh}}_{b}  = \mathbf{Y}^{\text{sh}}_{b} \mathbf{U}_i. \label{eq_admittance_bus} å
\end{IEEEeqnarray}

\subsubsection{Transformers  (\texttt{transformer})}
Transformers, which are different from voltage regulators due to their galvanic isolation, can have delta, wye and zigzag winding configurations, with two or more winding sets. The choice of configuration leads to a phase angle off-set between the primary and secondary (\emph{i.e.}, vector group), in multiples of 30$^\circ$. If so equipped, the neutral can be grounded. No-load losses can be significant, and are therefore modeled explicitly. PowerModelsDistribution implements the comprehensive approach to transformer modeling detailed in \cite{Dugan2004}, where n-winding transformers are decomposed into a set of simple idealized two-winding transformers and lossy branches.

A simple idealized transformer, index $t$, modeled with transformation matrix $\mathbf{T}_t$, is
\begin{IEEEeqnarray}{C}
    \mathbf{U}_i =  \mathbf{T}_t \mathbf{U}_j, \quad \mathbf{T}_t^\text{H} \mathbf{I}_{tij} +  \mathbf{I}_{tji} = 0. \label{eq_tf}
\end{IEEEeqnarray}
Transformers can also be equipped with on-load tap changers, which can be either operated jointly (gang operation), or separately for each phase.

\subsubsection{Loads, Generators, DER and Storage (\texttt{load, gen, storage})}
Power consumption and generation devices can be configured with wye or delta connections; phase-to-neutral configurations are special cases of wye connections.

In practice, instantaneous load power depends on the voltage magnitude; such voltage-dependent behavior is often described through ZIP or exponential load models.

While photovoltaic components are currently approximated as fixed generators in the internal data model where the maximum real and reactive power generation capability can vary over time, storage systems can be modeled using PowerModels' {\em multi-network} feature, which enables multi-period optimization models; storage systems are modeled generically in such a way as to be able to represent a variety of different types of storage, like batteries, fuel cells, or flywheels \footnote{https://lanl-ansi.github.io/PowerModels.jl/stable/storage}.

\subsubsection{Switches (\texttt{switch})}
Breakers, fuses, switches or sectionalizers allow for altering the topology of an existing network. Switches can be fixed open, fixed closed, or be optimized.

\subsection{Unbalanced OPF Formulations}
The unbalanced formulations currently implemented in PowerModelsDistribution are detailed in Table \ref{tab_formulations}, including their category (cat.), their variable space (var.), their coordinate space (coord.), their mathematical complexity (compl.), and their functional representation (repr.). These classifications are discussed in detail below. Note that each formulation collects mathematical formulations for all the components discussed, not solely for branches.

\begin{table}[tbh]
    \setlength{\tabcolsep}{1pt}
    \centering
    \caption{Features of unbalanced power flow formulations}\label{tab_formulations}
    \begin{tabular}{l l l l l l l }
        \hline
        Name & cat.& var. & coord. & compl. & repr.  & ref.\\
        \hline
        ACPPowerModel & BIM & SU & polar & NLP  & trig.   & \cite{Mahdad2006} \\
        ACRPowerModel & BIM & SU & rect. & NLP  & quadr.     &\\
        IVRPowerModel & BIM & IU & rect. & NLP & quadr. & \\
        SDPUBFPowerModel & BFM & SW & rect. & SDP  & conic   &  \cite{Gan2014}\\
        SOCConicUBFPowerModel & BFM & SW & rect. & SOC  & conic   &\cite{Gan2014,Kim2003} \\
        SOCNLPUBFPowerModel & BFM & SW & rect. & SOC  & quadr.  & \cite{Gan2014,Kim2003}  \\
        LPUBFDiagPowerModel & BFM & SW & rect.& LP  & lin.   &  \cite{Gan2014,Stewart2016,Dobbe2018}\\
        DCPPowerModel & BIM & SU & polar  & LP  & lin.   & \\
        \hline
    \end{tabular}
\end{table}
The upcoming subsections detail why and how these formulations are categorized.

\subsubsection{Category}
We can categorize formulations into either a bus injection model (BIM) or a branch flow model (BFM) \cite{Low2014}.

\subsubsection{Variable Space}
While the component models were illustrated in current-voltage vector variables,
the same physics can be represented in different variable spaces. This mechanism is used to obtain mathematical equations which have different behavior when solved with numerical methods. We categorize some recent work in mathematical formulations for unbalanced OPF depending on the choice of variable space for the branch flow equations into three distinct sets:
\begin{itemize}
    \item current $\mathbf{I}_{lij}$ - voltage $\mathbf{U}_i$,% (Kirchhoff's Laws);
    \item power $\mathbf{S}_{lij}$ - voltage $\mathbf{U}_i$,% (the `AC power flow equations' extended to unbalance \cite{Mahdad2006});
    \item power $\mathbf{S}_{lij}$ - lifted voltage $\mathbf{W}_{ij}= \mathbf{U}_i\mathbf{U}_j$.%, (as used in the convex relaxation formulations \cite{Gan2014}).
\end{itemize}

\subsubsection{Coordinates}
Either rectangular or polar coordinates can represent complex voltage variables.

\subsubsection{Mathematical Complexity}
We can have continuous convex and non-convex formulations of different complexities,
\begin{IEEEeqnarray}{C}
    \text{NLP} \supset \text{SDP} \supset \text{SOC} \supset \text{LP}.
\end{IEEEeqnarray}

\subsubsection{Representation}
We can distinguish quadratic and conic representations of convex forms. For example, given $ x_3\geq0, x_4 \geq 0$:
\begin{IEEEeqnarray}{C}
    (x_1)^2 + (x_2)^2 \leq x_3 x_4 \iff
    \left\Vert\begin{bmatrix}
    2 x_1   \\
    2 x_2    \\
    x_3 - x_4 \\
    \end {bmatrix}\right\Vert \leq x_3 + x_4. \nonumber
\end{IEEEeqnarray}
Note that linear constraints are both quadratic and conic and therefore only have one representation. The polynomial form can be used in conjunction with gradient-based solvers, whereas SOC solvers have norm-based interfaces. Trigonometric expressions, such as the AC polar form power flow equations, which uses sine and cosine, are not polynomial.

\section{Using PowerModelsDistribution} \label{sec_code_examples}
Following the design of PowerModels \cite{Coffrin2017}, each power system optimization problem (\emph{e.g.}, unbalanced OPF) has well-defined semantics for a large set of formulations (\emph{e.g.}, AC in polar coordinates, DC approximation, or SOC relaxation).

\subsection{Abstract Problem Definitions}
Several key problem definitions are already included in PowerModelsDistribution. In particular, Power Flow, Optimal Power Flow, and Maximal Load Delivery \cite{Coffrin2019}. Code Block \ref{code:jump_ex} shows a typical example of the OPF problem definition; first, variables for voltage, branch and transformer flows, generators and storage are initialized for the specified formulation, and then constraints are applied, including constraints for power balance, Ohm's Law, thermal limits, etc. Finally, a objective function, in this case a standard minimum fuel cost objective, is added to the problem. Details of each JuMP model will vary according to chosen formulation.

\begin{code}[t]
\caption{Problem specification for unbalanced OPF.}
\label{code:jump_ex}
\begin{minted}{julia}
function post_mc_opf(pm::AbstractPowerModel)
    variable_mc_voltage(pm)
    variable_mc_branch_flow(pm)
    variable_mc_transformer_flow(pm)
    variable_mc_generation(pm)
    variable_mc_storage(pm)

    constraint_mc_model_voltage(pm)

    for i in ids(pm, :ref_buses)
        constraint_mc_theta_ref(pm, i)
    end

    for i in ids(pm, :bus)
        constraint_mc_power_balance(pm, i)
    end

    for i in ids(pm, :storage)
        constraint_storage_state(pm, i)
        constraint_storage_complementarity_nl(pm, i)
        constraint_mc_storage_loss(pm, i)
        constraint_mc_storage_thermal_limit(pm, i)
    end

    for i in ids(pm, :branch)
        constraint_mc_ohms_yt_from(pm, i)
        constraint_mc_ohms_yt_to(pm, i)

        constraint_mc_voltage_angle_difference(pm, i)

        constraint_mc_thermal_limit_from(pm, i)
        constraint_mc_thermal_limit_to(pm, i)
    end

    for i in ids(pm, :transformer)
        constraint_mc_trans(pm, i)
    end

    objective_min_fuel_cost(pm)
end
\end{minted}
\end{code}

\subsection{Abstract Formulations}
Starting from a generic specification of an unbalanced OPF problem, PowerModels demonstrated that it is possible to specialize the model into concrete mathematical programs for given power flow formulations \cite{Coffrin2017}. Formulations define a representation of the electrical physics in a certain variable space (\emph{e.g.}, power-voltage), with a choice of coordinates (\emph{e.g.}, polar or rectangular). PowerModelsDistribution maintains the assumption that the combination of an abstract problem and a mathematical formulation results in a fully specified mathematical program, which is encoded as a JuMP model. Independently, the user can select a JuMP-compatible solver to solve the program.

\subsection{Data Formats}
During initial development we considered support for a variety of existing network data formats, ultimately settling on the OpenDSS format, which exhibits a good balance between its detail and inherent readability. In PowerModelsDistribution, only a subset of the OpenDSS format is supported; in particular, we support a subset of components for conversion into the internal data model, which follows closely that of PowerModels \cite{Coffrin2017}, including Loads, Capacitors, Reactors (only as shunts, not generically), Lines and Linecodes, Transformers, Generators, PVSystems, and Storage. It should also be noted that like PowerModels, and unlike OpenDSS, PowerModelsDistribution uses a non-dimensional unit system (per-unit) in its internal model to facilitate numerical stability.

In general, OpenDSS functions are not supported, with the notable exception of data-handling functions such as setting properties, and redirecting to additional files. That being said, the parser can ingest any valid OpenDSS files into a serializable data structure for additional user processing. Furthermore, several advanced input styles are fully supported, such as upper triangular matrices and reverse polish notation. Additional notes about support for the OpenDSS format can be found in the package documentation.\footnote{https://lanl-ansi.github.io/PowerModelsDistribution.jl}

\section{Proof-of-concept Study} \label{sec_study}
\subsection{Test Cases and Computational Setting}
This study considers 4 of the IEEE unbalanced test feeders, \emph{i.e.}, IEEE 13, 34, 123 and LVTestCase \cite{Mather2017}. We use the `.dss' case definitions, to facilitate comparison of power flow results w.r.t. OpenDSS. Comparing formulations is a multifaceted endeavor, requiring both high quality implementations of the mathematical models, as well as non-trivial test cases; the toolbox needs to facilitate comparisons of feasibility, optimality, computation time, memory usage and reliability (convergence).

In Table \ref{tab_test_case_features} we highlight the core features used in the IEEE distribution feeder test cases.  All cases are Kron-reduced and do not feature explicit neutrals. The table shows the types of branch shunts (branch sh.), noting whether they are full matrices, diagonal matrices, or contain none, the types of transformers (Wye-Wye (Yy), Delta-Wye (Dy), or Delta-Delta (Dd)), the types of loads (constant impedance (Z), constant current (I), constant power (P)), whether they contain bus shunts (bus sh.), and whether they have less than full three-phase branches ($<3$-p. branch).
\begin{table}[tbh]
    \setlength{\tabcolsep}{1pt}
    \centering
    \caption{Features used in IEEE test cases}\label{tab_test_case_features}
    \begin{tabular}{l c c c c c c c}
        \hline
        Case & branch sh. & transformer & loads & bus sh. & $<3$-p. branch & ref. \\
        \hline
        IEEE 13& diag & Yy, Dy & ZIP  & yes & yes & \cite{Kersting1991} \\
        IEEE 34& diag & Yy, Dy  & ZIP  & yes  & yes & \cite{Kersting1991} \\
        IEEE 123& full  & Yy, Dd & ZIP  & yes  & yes & \cite{Kersting1991}\\
        LVTestCase& none &Dy  & P  & no  & no & \cite{IEEEPES} \\
        \hline
    \end{tabular}
\end{table}

We include two studies: (i) a comparison of power flow results w.r.t. OpenDSS, using the AC polar form of the power flow equations, and (ii) a demonstration of the proposed abstract OPF problem evaluated using a representative set of formulations from the literature. The first requires support for full matrix branch shunts, transformers and ZIP loads, while the second uses simplified versions of the IEEE cases; transformers are converted to branches, and all loads are converted to wye-connected, constant-power loads. Furthermore, to obtain a non-dimensional voltage profile after simplification within $[0.9, 1.1]$, load set-points were reduced by 50\%. Because of the severe load reduction that would have been required to ensure feasibility after the simplification, IEEE123 was omitted from the simplified feeders. The results for AC formulations were obtained with Ipopt v3.12.10 as a solver, while SDP and SOC formulations utilized Mosek v9.0.95.

\subsection{Quality and Run-time Analysis}
\newcommand{\abs}[1]{\left|#1\right|}

The key metrics from the feasibility study are presented in Table \ref{tab_results_feasibility}.
We define the relative error $\delta$ as the largest relative difference in voltage magnitude across all network buses,
\begin{align}
	\delta = \max_{i,p}{\abs{ \frac{\abs{U}_{ip}^\text{OpenDSS}-\abs{U}_{ip}}{\abs{U}^\text{OpenDSS}_{ip}} }},
\end{align}
where $\abs{U}^\text{OpenDSS}_{ip}$ denotes the voltage magnitude at bus $i$ and phase $p$ obtained from OpenDSS. The only exception is in IEEE123, which contains a floating bus (610), where its phase-to-neutral voltages are only unique up to a constant. Therefore, at the floating bus, the phase-to-phase voltages are instead compared. As reported in Table \ref{tab_results_feasibility}, we note that the largest relative difference in voltage magnitude is 1.4e-7, indicating a close match in power flows w.r.t. OpenDSS.

\begin{table}[h!]
	\centering
	\caption{Comparison of power flow results of distribution test feeder cases w.r.t. OpenDSS}
	\label{tab_results_feasibility}
	\begin{tabular}{rlllll}
		\toprule
		&			&			&				&\multicolumn{2}{c}{LVTestCase}\\\cmidrule(r){5-6}
		                        & IEEE13 & IEEE34 & IEEE123 	& t=500 & t=1000\\\midrule
		$\delta$                & 5.1E-8 & 1.4E-7 & 1.3E-8 & 3.2E-8 & 3.3-8\\
		$\min{\abs{U}_{ip}}$    & 0.9750 & 0.9166 & 0.9858 & 1.0353 & 1.0226\\
		$\max{\abs{U}_{ip}}$    & 1.0686 & 1.0500 & 1.0437 & 1.0499 & 1.0496\\
		\bottomrule
	\end{tabular}
\end{table}

Formulation comparison results are presented in Table \ref{tab_results}. For each test case we also specify the number of nodes $|N|$ and the number of buses $|E|$. AC-NLP-polar and AC-NLP-rect are equivalent formulations, therefore yielding the same result. The SDP BFM and SOC BFM formulations are relaxations of the nonlinear BFM, equivalent to AC-NLP-polar; so their optimality gap w.r.t. AC-NLP-polar should be positive. For IEEE13, SDP BFM is tight up to the accuracy of the solvers, explaining the negative value. Since SOC BFM is a relaxation of SDP BFM, its optimality gap w.r.t. AC-NLP-polar should be the same or larger; for IEEE13, SDP BFM is tight whilst SOC BFM has a significant gap of 1.1\%. It should be noted that AC-NLP-rect and SOC BFM experienced numerical issues for IEEE13 and LVTestCase, respectively, illustrating the usefulness of the availability of multiple formulations.
\begin{table*}[tbh]
    \centering
    \caption{Optimality and runtime results for various unbalanced OPF formulations. LVTestCase evaluated at $t=1000$.}\label{tab_results}
    \begin{tabular}{l l l l l l l l l l l l}
        \hline
        &  &  & $\$/h$  & $\$/h$ & gap (\%)  & gap (\%)  & (s)  & (s) &(s)& (s) \\
        Test case & $|N|$ & $|E|$ & AC-NLP-polar  & AC-NLP-rect & SDP BFM  & SOC BFM  &  AC-NLP-polar  & AC-NLP-rect & SDP BFM & SOC BFM \\ \hline
        IEEE 13& 38 & 13  & 1.7598  & n.s. & -2.6E-3  & 1.1  & 0.0544 & n.s. & 0.2196 & 0.0819 \\
        IEEE 34& 135 & 34 & 0.9557  & 0.9557 & 0.47  & 1.8  & 0.9106 & 58.3162 & 0.6437 & 0.3317 \\
        LVTestCase & 2718 & 906 & 0.0243  & 0.0243 & 0.071  & n.s.  & 4.7184  & 34.3739 & 10.8802 & n.s. \\
        \hline
    \end{tabular}
\end{table*}

\section{Conclusions} \label{sec_conclusions}
This paper discusses the design and implementation of an open-source toolkit for simulation and optimization of distribution network power flows. This toolbox complements established tools such as OpenDSS and Gridlab-D by providing an optimization-first approach to distribution system modeling. Numerical results presented illustrate that the proposed approach provides comparable numerical results on a variety of unbalanced IEEE test feeders. We further demonstrate how modelers can use PowerModelsDistribution to explore exact models, relaxations and approximations in different variable spaces, and can interface with a wide array of state-of-the-art solvers through the Julia-based mathematical programming package, JuMP. PowerModelsDistribution presents a practical and well-tested domain-specific mathematical programming framework for distribution network optimization. We hope that this package can serve as a foundational tool enabling the speedy development of optimization models for distribution network problems, and, to that end, we encourage the research community to develop and contribute novel problem specifications and mathematical formulations. Furthermore, the extensible design of PowerModelsDistribution lends itself well to becoming a backend optimization tool such as PowerModels has been in \emph{e.g.}, PandaPower \footnote{https://github.com/e2nIEE/pandapower} and PowerSystems \footnote{https://github.com/NREL/PowerSystems.jl}.

The development of PowerModelsDistribution is an ongoing effort and in the near future we have plans to support additional high-priority features such as,
\begin{itemize}
    \item explicit representation of the neutral and/or ground (4-wire), and Kron-reduced representations \cite{Usman2020,Ciric2003};
    \item voltage regulators (\textit{i.e.}, auto-transformers) \cite{Bazrafshan2018a,Bazrafshan2019};
    \item short-circuit calculation;
    \item harmonics analysis \cite{Ying-YiHong2002};
    \item problem-specific solvers, \emph{e.g.}, power flow algorithms based on Newton-Raphson.
\end{itemize}

In addition, we hope to explore some experimental features, such as probing the feasibility of solving optimization problems in SI units rather than in the non-dimensional (per-unit) representation we currently utilize, and testing scalability for larger networks, \emph{e.g.}, urban distribution networks.

Finally, this work highlights a need for a library of unbalanced OPF benchmark cases that would enable easy comparison between the various distribution network solvers and simulators, such as the PGlib  \cite{Babaeinejadsarookolaee2019} AC-OPF benchmarks, which are only applicable to single-phase power flow equations.

\section{Acknowledgements}
This work was supported by funding from the U.S. Department of Energy’s (DOE) Office of Electricity (OE) as part of the CleanStart-DERMS project of the Grid Modernization Laboratory Consortium, and by the U.S. Department of Energy through the Los Alamos National Laboratory LDRD Program and the Center for Nonlinear Studies.

\end{document}